\documentclass[prd,preprint,preprintnumbers,showpacs,amsmath,amssymb,%
eqsecnum,nofootinbib]{revtex4}
\usepackage{graphicx}
\usepackage{float}
\usepackage{bm}
\begin{document}
\title{Asymptotic behavior of dynamical variables and naked
  singularity formation in spherically symmetric gravitational
  collapse}
%%%%%%%%%%%%%%
\author{Hayato Kawakami}
%\email{kawakami@gravity.phys.nagoya-u.ac.jp}
\author{Eiji Mitsuda}
\email{emitsuda@gravity.phys.nagoya-u.ac.jp}
\author{Yasusada Nambu}
\email{nambu@gravity.phys.nagoya-u.ac.jp}
\author{Akira Tomimatsu}
\email{atomi@gravity.phys.nagoya-u.ac.jp}
\affiliation{Department of Physics, Graduate School of Science, Nagoya
University, Nagoya 464-8602, Japan}
%%%
\date{June 23, 2009} % accepted version 1.0
%%%%%%%%%%%%%%%%
\begin{abstract}
  In considering the gravitational collapse of matter, it is an important
  problem to clarify what kind of conditions leads to the formation of
  naked singularity. For this purpose, we apply the 1+3 orthonormal
  frame formalism introduced by Uggla \textit{et al.} to the spherically
  symmetric gravitational collapse of a perfect fluid. This formalism
  allows us to construct an autonomous system of evolution and
  constraint equations for scale-invariant dynamical variables
  normalized by the volume expansion rate of the timelike orthonormal
  frame vector. We investigate the asymptotic evolution of such
  dynamical variables towards the formation of a central singularity
  and present a conjecture that the steep spatial gradient for the
  normalized density function is a characteristic of the naked
  singularity formation.
 \end{abstract}
\maketitle

%%%%%%%%%%%%%%%%%%%%%%%%%%%%%%%%%%%%%%%%%%%
\section{Introduction\label{sec:intro}}
The investigation of the end states of gravitational collapse of
sufficiently massive stars is a long standing problem to be pursued in
general relativity. One of the remarkable results is the naked
singularity formation and this is one of the possible end states in
spherically symmetric gravitational collapse starting from regular
initial data sets with nonzero measure. This was firstly shown by the
detailed analysis of the Lema\^itre-Toleman-Bondi (LTB) solution
describing the inhomogeneous dust gravitational collapse, and has been
subsequently confirmed in several other matter models including
perfect fluid models
\cite{SinghTP:CQG13:1996,JhinganS:CQG13:1996,HaradaT:PTP107:2002,HaradaT:P63:2004,DeshingkarSS:X07101866:2007,GiamboR:X08020992:2008}. In
relation to the cosmic censorship conjecture, much attention therefore
has been paid to the generality and the physical reasonability of the
initial data sets leading to the naked singularity formation in the
various matter models (see \cite{ HaradaT:PRD58:1998,
  JhinganS:PRD61:2000, GoncalvesSMCV:PRD65:2002, JoshiPS:PRD69:2004,
  MahajanA:CQG22:2005} for example).

Besides the initial data analysis, dynamical aspects of  spherically
symmetric gravitational collapse at later stages have been studied in
terms of the shear function associated with the velocity vectors of
the collapsing matter near the center \cite{JoshiPS:PRD65:2002,
  JoshiPS:PRD70:2004, MenaFC:PRD70:2004}, and it has been suggested
that the shear can contribute to the delay of the apparent horizon
formation. This role of the shear function may be important in
considering the dynamical mechanism of the naked singularity
formation. In this paper, however, we would like to focus on another
dynamical approach to the end-state problem in  spherically
symmetric gravitational collapse. Our main concern is the relationship
between asymptotic behavior of dynamical variables towards the
formation of a central singularity and the causal structure of the
arising singularity.

As an useful method to analyze such dynamical properties, we adhere to
the 1+3 orthonormal frame formalism which was originally proposed by
Elst and Uggla \textit{et al.}~\cite{vanElstH:CQG14:1997,
  UgglaC:PRD68:2003} to study the dynamical behavior of the
gravitational field variables such as the shear $\sigma_{ab}$ and the
expansion $\Theta$ associated with the timelike orthonormal frame
vectors near the spatially inhomogeneous cosmological initial
singularity. This formalism is based on the coordinate independent
representation of Einstein's field equations in the form of an
autonomous system of the first order evolution equations and
constraints with the scale-invariant dimensionless variables
normalized by the Hubble scalar $H\equiv \Theta/3$.  By virtue of this
Hubble normalization of dynamical variables, it is possible to remove
the time dependent factors due to the volume contraction given by the
rate $\Theta$ measured in the local reference frame. As will be shown
in this paper, the Hubble normalized density function $\Omega$ can
approach zero or remain finite with the lapse of time towards
singularity formation in the spherically symmetric inhomogeneous
perfect fluid collapse, even though the central proper density becomes
divergent at this final stage.

Our main purpose in this paper is to discuss the relation between the
asymptotic behavior of $\Omega$ and the causal structure of the
singularity. For the LTB solution, using the causal structure
classified by the papers\cite{SinghTP:CQG13:1996,JhinganS:CQG13:1996},
we find that the growth of the steep spatial gradient of the profile
of $\Omega$ near the center in the case $\Omega\rightarrow 0$ is a
characteristic property leading to the naked singularity formation. If
the density function $\Omega$ remains finite at the final stage, the
end-state problem becomes more subtle. Nevertheless, we can discuss
the critical value of the density contrast which gives a threshold of
the transition from the black hole formation to the naked singularity
formation.

This paper is organized as follows: In Sec.~\ref{sec:EC}, we begin
with a brief review of the 1+3 orthonormal frame formalism with the
Hubble normalized variables and apply it to a spherically symmetric
perfect fluid system. In this formalism, we adopt the separable volume
gauge, which specifies the lapse function to be equal to the inverse
of the Hubble scalar $H$. In Sec.~\ref{sec:LTB}, considering the
inhomogeneous dust gravitational collapse described by the marginally
bound LTB solution, we show that this gauge condition is useful to
examine the asymptotic behavior of the Hubble normalized variables
towards the central singularity formation. Then, we relate the
asymptotic behavior of the Hubble normalized density function $\Omega$
to the arising causal structure of the end states of collapse. We
propose a conjecture that the larger spatial gradient of the
asymptotic profile of $\Omega$ is essential to the naked singularity
formation. In Sec.~\ref{sec:PF}, to support this conjecture, the
asymptotic analysis is extended to gravitational collapse of perfect
fluid with pressure.  Taking account of the causal structure of
spherically symmetric self-similar spacetimes which has been clarified
by previous works\cite{OriA:PRD42:1990, CarrBJ:PRD62:2000,
  CarrBJ:PRD67:2003}, we numerically estimate the critical value of
the density contrast using the asymptotic profile of $\Omega$ for the
naked singularity formation. The results are summarized in
Sec.~\ref{sec:sum}. Throughout this paper, the units in which $8\pi
G=c=1$ are used.

%%%%%%%%%%%%%%%%%%%%%%%%%%%%%%%%%%%%%%%%%%%%%%%%%%%%%%%%%%%%%%%%%%%%%%%%%%%%%%%%%
\section{Basic equations  for Hubble normalized variables \label{sec:EC}}

In this section, following the 1+3 orthonormal frame formalism
developed in \cite{vanElstH:CQG14:1997, UgglaC:PRD68:2003}, we present
an autonomous system of evolution equations and the constrains for the
scale-invariant variables in a spherically symmetric system with a
perfect fluid source. We express an orthonormal frame as
$\{\boldsymbol{e}_{0},\boldsymbol{e}_{\alpha}\}$ (where
$\alpha=1,2,3$) with the unit vectors $\boldsymbol{e}_{0}$ and
$\boldsymbol{e}_{\alpha}$ representing the timelike reference
congruence and the rest of 3-spaces, respectively. The frame metric is
given by $\eta_{\mu\nu}=\text{diag}[-1,1,1,1]$. For simplicity, we
specify the timelike frame vector $\boldsymbol{e}_{0}$ to be
hypersurface orthogonal, and the spacelike
frame vectors $\boldsymbol{e}_{\alpha}$ to be nonrotating
Fermi-propagated along the integral curves of $\boldsymbol{e}_{0}$.

Defining the four-velocity $\boldsymbol{u}\equiv\boldsymbol{e}_{0}$
for the unit timelike vector $\boldsymbol{e}_{0}$ tangent to the
reference congruence, we can introduce the basic geometrical
quantities $(\dot{u}^{\alpha},\Theta, \sigma^{\alpha\beta},a^{\alpha},
n^{\alpha\beta})$ through the commutator relations as follows
%%%%
\begin{align}
 &[\boldsymbol{e}_{0},\boldsymbol{e}_{\alpha}]=\dot{u}_{\alpha}\boldsymbol{e}_{0}
-\left(\frac{\Theta}{3}\delta_{\alpha}^{\ \beta}+\sigma_{\alpha}^{\ \beta}\right)
 \boldsymbol{e}_{\beta}~,\label{eq:com}\\
 &[\boldsymbol{e}_{\alpha},\boldsymbol{e}_{\beta}]=
\left(2a_{[\alpha}\delta_{\beta]}^{\ \gamma}
 +\epsilon_{\alpha\beta\delta}n^{\delta\gamma}\right)\boldsymbol{e}_{\gamma}~,
\end{align}
%%%
where the square brackets denote the antisymmetric part of a tensor,
and $\epsilon_{\alpha\beta\gamma}$ is the totally antisymmetric three
dimensional permutation tensor. The scalar function $\Theta$ is the
volume expansion rate, the vector $\dot{u}^{\alpha}$ and the tensor
$\sigma^{\alpha\beta}$ (the trace-free symmetric tensor) are the
acceleration rate and the shear rate of the frame vector
$\boldsymbol{e}_{0}$, respectively. They are calculated from the
equations
%%%%
\begin{align}
 &\Theta = \nabla_{\mu}u^{\mu}~,\\
 &\dot{u}_{\alpha} = u^{\mu}\nabla_{\mu}u_{\alpha}~,\\
 &\sigma_{\alpha\beta} = \nabla_{(\beta}u_{\alpha)}
-\frac{\Theta}{3}\left(\eta_{\alpha\beta}+u_{\alpha}u_{\beta}\right)
+\dot{u}_{(\alpha}u_{\beta)}~,
\end{align}
%%%
where the round brackets denote the symmetric part of a tensor. (See
\cite{vanElstH:CQG14:1997} for the definition of the covariant
derivative in the orthonormal frame formalism.) In addition, the
quantities $a^{\alpha}$ and $n^{\alpha\beta}$ (the symmetric tensor)
determine the connection on the one-parameter family spacelike
hypersurfaces which can be defined by the assumption of the
hypersurface-orthogonality of the timelike frame vector
$\boldsymbol{e}_{0}$.

Now let us turn our attention to the introduction of the basic
variables characterizing the matter field.  We consider a perfect
fluid  as the collapsing matter in this paper. We assume the
equation of state to be
%%%
\begin{equation}
 \tilde{p} = (\gamma-1)\tilde{\mu} \label{eq:state}
\end{equation}
%%%
with the constant $\gamma$ lying in the range $1\leq \gamma \leq 2$.
The pressure $\tilde{p}$ and the energy density $\tilde{\mu}$ are
measured by a comoving observer who has the same velocity as the fluid
four-velocity $\tilde{\boldsymbol{u}}$. In general, the fluid
four-velocity $\tilde{\boldsymbol{u}}$ is not equal to the
four-velocity $\boldsymbol{u}$ defined by the timelike frame vector
$\boldsymbol{e}_{0}$, thus we introduce the basic matter variables
$\mu$ and $\boldsymbol{v}$ by decomposing the energy-momentum tensor
of the perfect fluid with respect to $\boldsymbol{u}$ into the form
%%%
\begin{equation}
 T_{\mu\nu}=\mu\left\{u_{\mu}u_{\nu}+\gamma
   G^{-1}\left(2v_{(\mu}v_{\nu)}
+v_{<\mu}v_{\nu>}\right)\right\}+p\left(g_{\mu\nu}+u_{\mu}u_{\nu}\right)~,
\end{equation}
%%%
where the angle brackets denote the symmetric trace-free part of a
tensor. We have the following relations
%%%
\begin{equation}
 \mu = \Gamma^{2}G\tilde{\mu}~, \qquad p = G^{-1}\left\{\gamma-1
+\left(1-\frac{2}{3}\gamma\right)v_{\mu}v^{\mu}\right\}\mu
\end{equation}
%%%
with the scalar functions $G$ and $\Gamma$ are defined by
%%%
\begin{equation}
G \equiv 1 + (\gamma-1)v_{\mu}v^{\mu}~, \qquad \Gamma \equiv
 \frac{1}{\sqrt{1-v_{\mu}v^{\mu}}}~.
\end{equation}
%%%
The vector $\boldsymbol{v}$ represents the peculiar fluid velocity
relative to the rest 3-spaces of $\boldsymbol{e}_{0}$, and defined
through the relations
%%%
\begin{equation}
 \tilde{u}^{\mu} \equiv \Gamma\left(u^{\mu}+v^{\mu}\right)~, \qquad u_{\mu}v^{\mu}=0
\end{equation}
%%%
with the fluid four-velocity $\boldsymbol{\tilde{u}}$ normalized as
$\tilde{u}_{\mu}\tilde{u}^{\mu}=-1$.
 
An important procedure of the formalism developed in
\cite{UgglaC:PRD68:2003} is to introduce the scale-invariant
 dimensionless variables by normalizing the geometrical and matter
variables using the Hubble scalar
%%%
\begin{equation}
 H \equiv \frac{\Theta}{3}~.
\end{equation}
%%%
We denote the Hubble normalized quantities as
%%%
\begin{align}
 &\boldsymbol{\partial}_{0} \equiv \frac{\boldsymbol{e}_{0}}{H}~, \qquad
 \boldsymbol{\partial}_{\alpha} \equiv
 \frac{\boldsymbol{e}_{\alpha}}{H}~,\label{eq:HN1}\\
 &\left\{\dot{U}^{\alpha},\Sigma_{\alpha\beta},A^{\alpha},N^{\alpha\beta}\right\}\equiv
 \frac{1}{H}\left\{\dot{u}^{\alpha},\sigma_{\alpha\beta},
   a^{\alpha},n^{\alpha\beta}\right\}~,\label{eq:HN2} \\
 &\Omega \equiv \frac{\mu}{3H^{2}}~,\label{eq:HN3}
\end{align}
%%%
where $\Omega$ is the Hubble normalized density function.  The vector
$\boldsymbol{v}$ is a dimensionless variable and the Hubble
normalization is not necessary for this variable. To introduce a local
coordinate system, we also define the Hubble normalized components of
the frame vectors as
%%%
\begin{equation}
 E_{\mu}^{\ \hat{a}} \equiv \frac{e_{\mu}^{\ \hat{a}}}{H}~.\label{eq:HN4}
\end{equation}
%%%
As expressed in Eq.~\eqref{eq:HN4}, we hereafter attach the hat
$\hat{}$ to spacetime indices in order to distinguish them from the
orthonormal frame indices.  As physically interesting additional
scale-invariant variables, we introduce the deceleration scalar $q$
and the spatial Hubble gradient $\lambda_{\alpha}$ defined by
%%%
\begin{align}
 &q \equiv -1-\frac{1}{H}\boldsymbol{\partial}_{0}H~,\label{eq:HN5}\\
 &\lambda_{\alpha}\equiv -\frac{1}{H}\boldsymbol{\partial}_{\alpha}H~.\label{eq:HN6}
\end{align}
%%%
They will be used to eliminate the Hubble scalar $H$ appearing in the
evolution equations and the constraints.\footnote{Although the
  spatial Hubble gradient was expressed as $r_{\alpha}$ in
  \cite{UgglaC:PRD68:2003}, we have changed its notation in order to
  prevent readers from confusing it with the radial coordinate $r$
  which will be introduced later.}

Some gauge choice is still allowed within the framework of the 1+3
orthonormal frame formalism, and in \cite{vanElstH:CQG14:1997,
  UgglaC:PRD68:2003}, the evolution equations and the constraints for
the Hubble normalized variables have been given with the so-called
separable volume gauge, which simplifies the temporal frame derivative
$\boldsymbol{\partial}_{0}$ to
%%%
\begin{equation}
\boldsymbol{\partial}_{0}=-\partial_{t}\label{eq:gauge1}
\end{equation}
%%%
using a nondimensional time coordinate $t$.%
\footnote{While the lapse function is the positive definite function
  equal to the Hubble scalar $H$ in the separable volume gauge of
  \cite{UgglaC:PRD68:2003}, in this paper we specify the lapse
  function to be $-H$ in order to keep the positivity of the lapse
  function.  The minus sign in the right hand side of
  Eq.~\eqref{eq:gauge1}, which does not appear in the corresponding
  equation in \cite{UgglaC:PRD68:2003}, comes from this gauge
  specification, which may be referred to as the separable volume
  gauge for gravitational collapse.}%
 ~Further, the additional gauge constraint
%%%
\begin{equation}
 \dot{U}_{\alpha}=\lambda_{\alpha}
\end{equation}
%%%
is required for the separable volume gauge (see
\cite{UgglaC:PRD68:2003} for its details). Owing to this specification
of the gauge, the matching of the four-velocity $\boldsymbol{u}=
\boldsymbol{e}_{0}$ of the reference congruence with the fluid
four-velocity $\boldsymbol{\tilde{u}}$ (i.e.,
$\boldsymbol{v}=\boldsymbol{0}$) as in \cite{SussmanRA:CQG25:2008} is
not always permitted. The important geometrical result of
Eq.~\eqref{eq:gauge1} with the commutator equation \eqref{eq:com} is
that the volume density $\mathcal{V}$ defined by
$\mathcal{V}^{-1}\equiv \det(e_{\alpha}^{\ i})$ has the form
%%%
\begin{equation}
 \mathcal{V}=\mathcal{V}_0 \times e^{-3t},\label{eq:V}
\end{equation}
%%%
where $\mathcal{V}_0$ is an arbitrary function of spatial
coordinates. From Eq.~\eqref{eq:V}, in the limit $t\rightarrow\infty$,
the volume density approaches zero to form a singularity. This is a
useful property of the separable volume gauge to investigate the
asymptotic dynamical behavior just before the singularity formation
in gravitational collapse.

Now let us study spherically symmetric gravitational collapse with
perfect fluid using the Hubble normalized scale-invariant
variables. By virtue of the Hubble normalization, the Hubble scalar
$H$ becomes the only variable carrying a physical reference scale and
the analysis of the evolution equations and the constraints for these
variables will allow us to observe dynamical behavior deviated from
the time dependence of the volume contraction of the reference
congruence.  We consider the spherically symmetric line element of the
form
%%%
\begin{equation}
 ds^{2}=-I^{2}(t,r)dt^{2}+J^{2}(t,r)dr^{2}+R^{2}(t,r)\left(d\theta^{2}+
 \sin^{2}\theta d\phi^{2}\right)~. \label{eq:LE}
\end{equation}
%%%
The coordinate expressions for the orthonormal frame
derivatives can be given by
%%%
\begin{equation}
 \boldsymbol{e}_{0} = I^{-1}\partial_{\hat{t}}~, \qquad
\boldsymbol{e}_{r} = J^{-1}\partial_{\hat{r}}~, \qquad 
\boldsymbol{e}_{\theta} = R^{-1}\partial_{\hat{\theta}}~, \qquad
\boldsymbol{e}_{\phi} = R^{-1}\sin^{-1}\theta\partial_{\hat{\phi}}~. \label{eq:ofv}
\end{equation}
%%%
As is used in Eq.~\eqref{eq:ofv}, hereafter we substitute the letters
$r,\theta,\phi$ (instead of the numbers 1,2,3) into the spacetime
spatial indices $\hat{i}$ and the orthonormal frame spatial indices
$\alpha$ (i.e., $i=\hat{r},\hat{\theta},\hat{\phi}$ and
$\alpha=r,\theta,\phi$). The gauge conditions \eqref{eq:gauge1} and
\eqref{eq:V} are reduced to
%%%
\begin{equation}
 I = -H^{-1} \label{eq:gauge}
\end{equation}
%%%
and
%%%
\begin{equation}
 J(t,r) = C(r)R^{-2}e^{-3t}~,\label{eq:J}
\end{equation}
%%%
where $C$ is an arbitrary function of $r$.

An autonomous system of evolution equations and constraints for the
Hubble normalized variables presented in \cite{vanElstH:CQG14:1997,
  UgglaC:PRD68:2003} has been expressed with the orthonormal frame
derivatives applicable to any spacetimes. It is a straightforward to
apply the formalism to the spherically symmetric metric \eqref{eq:LE}
with the separable volume gauge. The quantities $(E_{r}^{\ \hat{r}},
E_{\theta}^{\ \hat{\theta}}, A^{r}, \lambda^{r}, \Sigma^{rr}, \Omega,
v^{r}, q)$ are independent Hubble normalized variables to be analyzed
here and we arrive at the following evolution equations for these variables
%%%
\begin{align}
&\dot{E}_{r}^{\ \hat{r}}=-\left(q-\Sigma^{rr}\right)E_{r}^{\ \hat{r}}~,\label{eq:EE1}\\
&\dot{E}_{\theta}^{\
  \hat{\theta}}=-\left(q+\frac{1}{2}\Sigma^{rr}\right)
E_{\theta}^{\ \hat{\theta}}~,\label{eq:EE2}\\
&\dot{A}^{r}=-\left(q-\Sigma^{rr}\right)A^{r}
-\frac{1}{2}E_{r}^{\ \hat{r}}(\Sigma^{rr})'~,\label{eq:EE3}\\
&\dot{\lambda}^{r}=-\left(q-\Sigma^{rr}\right)\lambda^{r}
-E_{r}^{\ \hat{r}}q'~,\label{eq:EE4}\\
&\dot{\Sigma}^{rr}=-(q-2)\Sigma^{rr}+\frac{2}{3}E_{r}^{\
  \hat{r}}(A^{r})'
-\frac{2}{3}E_{r}^{\
  \hat{r}}(\lambda^{r})'-\frac{4}{3}\lambda^{r}A^{r}
-\frac{2}{3}(E_{\theta}^{\ \hat{\theta}})^{2}
-2\gamma G^{-1}\Omega(v^{r})^{2}~,\label{eq:EE5}\\
&\dot{\Omega}=-(2q-1)\Omega+3G^{-1}\left\{\gamma-1
+\left(1-\frac{2}{3}\gamma\right)(v^{r})^{2}\right\}\Omega \label{eq:EE6}\\
&\qquad\qquad\qquad
 +\gamma E_{r}^{\ \hat{r}}(G^{-1}\Omega v^{r})'
+\gamma G^{-1}\Omega v^{r}\left(v^{r}\Sigma^{rr}-2A^{r}\right)~, \notag
\end{align}
%%%
and the constraints 
%%%
\begin{align}
&1+\frac{1}{3}\left\{2E_{r}^{\
    \hat{r}}(A^{r})'-2\lambda^{r}A^{r}-3(A^{r})^{2}
+(E_{\theta}^{\
  \hat{\theta}})^{2}\right\}-\frac{1}{4}(\Sigma^{rr})^{2}-\Omega=0~,
\label{eq:C1}\\
&E_{r}^{\ \hat{r}}(\Sigma^{rr})'+2\lambda^{r}-\Sigma^{rr}\lambda^{r}
-3A^{r}\Sigma^{rr}+3\gamma G^{-1}\Omega v^{r}=0~,\label{eq:C2}\\
&E_{r}^{\ \hat{r}}(E_{\theta}^{\ \hat{\theta}})'-(A^{r}
+\lambda^{r})E_{\theta}^{\ \hat{\theta}}=0~,\label{eq:C3}
\end{align}
%%%
where the dot and the prime mean the partial derivatives $\partial_{t}$
and $\partial_{r}$, respectively. These equations correspond to the
Einstein equations, the Jacobi identities and the contracted Bianchi
identities. In addition to these equations, we can use the following Raychaudhuri
equation for the deceleration scalar $q$:
%%% 
\begin{equation}
 q=\frac{1}{2}(\Sigma^{rr})^{2}-\frac{1}{3}E_{r}^{\
   r}\partial_{r}\lambda^{r}
+\frac{2}{3}\lambda^{r}A^{r}+\frac{1}{2}\left[1+3G^{-1}\left\{\gamma
    -1
+\left( 1-\frac{2}{3}\gamma\right) (v^{r})^{2}\right\}\right]\Omega~.\label{eq:q}
\end{equation}
%%%
It is remarkable that no time derivative of the radial velocity
$v^{r}$ appears in these set of the evolution equations and the
constraint equation \eqref{eq:C2} can be used to determine $v^{r}$. We
can also check that the remaining two constraints \eqref{eq:C1} and
\eqref{eq:C3} are consistent with other six evolution equations for
$(E_{r}^{\ \hat{r}}, E_{\theta}^{\ \hat{\theta}}, A^{r}, \lambda^{r},
\Sigma^{rr}, \Omega)$.

Finally, let us explicitly present the Hubble normalized
variables using the metric functions and the matter fields:
%%%
\begin{align}
&E_{r}^{\ \hat{r}}=-\frac{I}{J}~, \qquad E_{\theta}^{\
  \hat{\theta}}=-\frac{I}{R}~,\label{eq:g-matter1} \\
&\Sigma^{rr}=-\frac{2}{3}\left(\frac{\dot{J}}{J}-\frac{\dot{R}}{R}\right)~,
\label{eq:g-matter2} \\
&A^{r}=\frac{I\partial _{r}R}{JR}~,\label{eq:g-matter3} \\
&\lambda^{r}=-\frac{\partial _{r}I}{J}~, \qquad 
q=-1-\frac{\dot{I}}{I}~,\label{eq:g-matter4} \\
 &v^{r}=\frac{\tilde{u}^{r}}{\tilde{u}^{t}}~,\label{eq:g-matter6}\\
&\Omega =\frac{\left\{1+(\gamma-1)(v^{r})^{2}\right\}
I^{2}\tilde{\mu}}{3\left\{1-(v^{r})^{2}\right\}}~.\label{eq:g-matter5}
\end{align}
%%%
These relations are useful to analyze the asymptotic behavior of
dynamical variables. Although the description of time evolution with
the Hubble normalized variables and the separable volume gauge is
available until the singularity is formed at $t=\infty$, the spacetime
region covered by the radial coordinate $r$ may be too restricted to
determine the nakedness of the arising singularity.  Thus we will
restrict our investigations to the models of gravitational collapse of
which causal structure of the end states is already known. An example
of such a model is the LTB solution, which is well known as a generic
model of the inhomogeneous dust gravitational collapse. In the next
section, we will examine the dynamical behavior of the Hubble
normalized variables in the marginally bound LTB spacetime to
discuss their key feature relevant to the naked singularity
formation.

%%%%%%%%%%%%%%%%%%%%%%%%%%%%%%%%%%%%%%%%%%%%%%%%%%%%%%%%%%%%%%%%%%%%%%%%%%%%%%%%%%%%%%
\section{Asymptotic behavior in inhomogeneous dust collapse \label{sec:LTB}}

\subsection{Brief review of the LTB solution}
Let us begin with a brief review of the LTB solution describing
inhomogeneous dust gravitational collapse (see
\cite{SinghTP:CQG13:1996,JhinganS:CQG13:1996,HaradaT:PTP107:2002} for its details). This
solution includes the two arbitrary functions of radial coordinate
usually denoted as $F$ and $f$ , which are related to the Misner-Sharp
mass and the initial velocity, respectively. In this paper, we
consider only the solution with $f=0$, which is called the marginally
bound solution. Using the comoving coordinate system
$\{\tau,\rho,\theta,\phi\}$, the line element for this solution is
%%%
\begin{align}
 &ds^{2} = -\frac{4}{9B}d\tau^{2}+(\partial_{\rho}R)^{2}d\rho^{2}+R^{2}\left(d\theta^{2}
 +\sin^{2}\theta d\phi^{2}\right) , \\
%%%
 &R(\tau, \rho) = \rho\left\{1-\sqrt{\frac{F(\rho)}{B\rho^{3}}}(1+\tau)\right\}^{2/3}~,
\end{align}
%%%
where $B$ is an arbitrary positive constant.

The time $\tau=-1$ is usually interpreted as the initial time at which
the area radius $R$ becomes equal to the coordinate radius $\rho$. In
addition, without loss of generality, we can choose $\tau=0$ to be the
time of the central singularity formation by specifying the leading
order term of the arbitrary function $F(\rho)$ near the regular center
$\rho=0$ as $F\simeq B\rho^{3}$. In the comoving coordinate system,
the components of the dust four-velocity are given by
%%%
\begin{equation}
 \tilde{u}^{\hat{\tau}}= -\frac{3\sqrt{B}}{2}~, \qquad 
\tilde{u}^{\hat{\rho}}= \tilde{u}^{\hat{\theta}}=\tilde{u}^{\hat{\phi}}=0~,\label{eq:tildeu}
\end{equation}
%%%
and from the Einstein equations we obtain the proper energy density
$\tilde{\mu}$ measured by a comoving observer as
%%%
\begin{equation}
 \tilde{\mu}(\tau,\rho) = \frac{\partial_{\rho}F}{R^{2}\partial_{\rho}R}~.\label{eq:tildemu}
\end{equation}
%%%

The approximate form of the function $F$ near the regular center
$\rho=0$ can be written as
%%%
\begin{equation}
 F(\rho) = B\rho^{3}\left(1-2F_{n}\rho^{n}+O(\rho^{n+1})\right) \label{eq:F}
\end{equation}
%%%
with a positive integer $n$. The subleading term $F_n\rho^n$ in
Eq~\eqref{eq:F} represents the dominant inhomogeneity of an initial
dust distribution near the center, and we assume $n\geq 2$ and $F_n>0$
to require the regularity of the proper energy density at the center
(i.e., $\partial_{\rho}\tilde{\mu}=0$ at $\rho=0$ and at $\tau=-1$).
The causal structure of the end states of the collapse is closely
related to the inhomogeneity given by the term
$F_n\rho^n$. For $n=2$, the shell-focusing naked singularity
appears at the center $\rho=0$ at the time $\tau=0$. For $n\geq
4$, the arising singularity is hidden behind the event horizon. For
$n=3$, using the parameter
%%%
\begin{equation}
\label{eq:b}
 b\equiv \frac{F_{3}}{B^{3/2}} ,
\end{equation}
%%%
the condition for the naked singularity formation is given by
\cite{SinghTP:CQG13:1996,JhinganS:CQG13:1996}
%%%
\begin{equation}
\label{eq:bc}
b>b_c,\qquad  b_{c}=\frac{26+15\sqrt{3}}{4} .
\end{equation}
%%%

%%%%%%%%%%%%%%%%%%%%%%%%%%%%%%%%%%%%%%%%%%%%%%%%%%%%%%%%%%%%%%%%%%%%%%
\subsection{Asymptotic behavior of the Hubble normalized variables}

\begin{figure}[H]
\centering
\includegraphics[width=0.5\linewidth,clip]{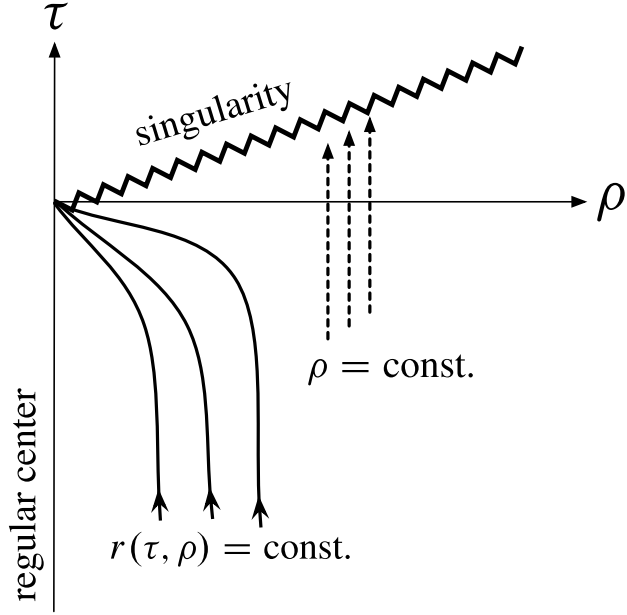}
\caption{A schematic diagram describing the relation between the
  comoving coordinate systems $\{\tau, \rho\}$ and the separable
  volume gauge coordinates $\{t,r\}$. In the limit
  $t\rightarrow\infty$, all the timelike curves labeled $r=$
  const. (solid curves) converge to the onset point $\tau=\rho=0$ of
  the singularity formation.} %describing $r=\text{const.}$ lines .}
\label{fig:diagram}
\end{figure}

In this subsection, we would like to clarify how the asymptotic
behavior of the Hubble normalized variables leading to the naked
singularity formation depends on the initial density inhomogeneity
characterized by the integer $n$. With the help of the relation
between the causal structure of the end states and the choice of $n$
obtained in \cite{SinghTP:CQG13:1996,JhinganS:CQG13:1996}, we discuss
what asymptotic behavior of the Hubble normalized variables
characterize the causal structure of the singularity. As the formation
of the central singularity occurs at the point $\tau=\rho=0$ in the
comoving coordinate system $\{\tau,\rho,\theta,\phi\}$, our strategy
is to analyze the asymptotic $t$-dependence of the Hubble normalized
variables in the limit $t\rightarrow\infty$ by using the coordinate
system $\{t,r,\theta,\phi\}$ with the separable volume gauge
condition.  For this purpose, we consider the coordinate
transformation between the two coordinate systems, which leads to the
following partial differential equations for $\tau(t,r)$, $\rho(t,r)$,
$I(t,r)$ and $J(t,r)$
%%%
\begin{align}
 &\frac{4}{9B}\dot{\tau}^{2}-(\partial_{\rho}R)^{2}\dot{\rho}^{2}=I^{2}~,\label{eq:CT1}\\
 &(\partial_{\rho}R)^{2}\rho^{\prime 2}-\frac{4}{9B}\tau^{\prime
   2}=J^{2}~,\label{eq:CT2} \\
 &(\partial_{\rho}R)^{2}\dot{\rho}\rho'=\frac{4}{9B}\dot{\tau}\tau^{\prime}~.\label{eq:CT3}
\end{align}
%%%
We also have Eq.~\eqref{eq:J} as the separable volume gauge condition.
As will be shown, the four Eqs. \eqref{eq:CT1}-\eqref{eq:CT3}
with \eqref{eq:J} demand that timelike curves with
$r=\text{const.}$ to converge to the singular point $\tau=\rho=0$. The
behavior of these coordinates is schematically shown in
Fig.~\ref{fig:diagram}.
%%%
%%%%%
\noindent
In particular, the exponential $t$-dependence of the function $JR^{2}$
in Eq.~\eqref{eq:J} significantly affects the asymptotic relation
between the two coordinate systems $\{\tau, \rho\}$ and $\{t,r\}$, and
the converging behavior of the $r=\text{const.}$ timelike curves is
useful to analyze the asymptotic dynamical features just before the
central singularity formation.  In this subsection, we analyze the
asymptotic $t$-dependence of the Hubble normalized variables on such a
congruence of the timelike curves.  When the coordinate variables
$\tau(t,r)$ and $\rho(t,r)$ approach zero in the limit
$t\rightarrow\infty$ with a fixed value of $r$, the metric function
$R(t,r)$ also goes to zero as
%%%
\begin{equation}
 R(\tau,\rho) \simeq \rho(F_{n}\rho^{n}-\tau)^{2/3}~. \label{eq:appR}
\end{equation}
%%%
The key issue to be analyzed here is which of the terms $F_n\rho^n$ and
$\tau$ in Eq.~\eqref{eq:appR} becomes dominant in the limit
$t\rightarrow\infty$ along the timelike curves.

From Eq.~\eqref{eq:J}, we assume that the metric functions $J$ and $R$
have exponential $t$-dependence in the limit
$t\rightarrow\infty$. This assumption turns out to be compatible with
Eqs.~\eqref{eq:CT1}-\eqref{eq:CT3} under the relations 
%%%
\begin{equation}
\label{eq:RIJ}
R\sim J\sim
\tau\sim I\sim \exp(-t), 
\end{equation}
%%%
if the exponential $t$-dependence of $\rho$ is also derived from
Eq.~\eqref{eq:appR}. To check this, we first assume that the ratio
$F_n\rho^n/|\tau|$ approaches zero. In this case, Eq.~\eqref{eq:appR}
leads to $\rho\sim\exp(-t/3)$, which is consistent with the assumption
$F_n\rho^n/|\tau|\rightarrow 0$ only for $n\geq 4$.  If the ratio
$F_n\rho^n/|\tau|$ is assumed to blow up, we have
$\rho\sim\exp\{-3t/(3+2n)\}$. It is easy to check that this is allowed
only for $n=2$, which gives the exponential $t$-dependence as
$\rho\sim\exp(-3t/7)$. For the remaining case $n=3$, we obtain
$\rho\sim\exp(-t/3)$ and the ratio $F_n\rho^n/|\tau|$ remains finite.

Now let us see the asymptotic behavior of the Hubble normalized
variables using the asymptotic exponential forms of the metric
functions $I$, $J$, $R$ and the coordinate variables $\tau$,
$\rho$. Because we have the relation Eq.~\eqref{eq:RIJ} irrespective
of the choice of $n$, Eqs.~\eqref{eq:g-matter1}-\eqref{eq:g-matter4}
mean that the Hubble normalized variables $(E_{r}^{\ \hat{r}},
E_{\theta}^{\ \hat{\theta}}, A^{r}, \lambda^{r})$ become finite and
their values can only depend on $r$.  On the other hand, variables
$\Sigma^{rr}$ and $q$ approach zero.  Using the condition
$\tilde{u}^{\hat{\rho}}=0$ for the comoving coordinate system, the
radial velocity $v^r$ given by \eqref{eq:g-matter6} can be rewritten
into the form
%%%
\begin{equation}
 (v^r)^2=\frac{J^2\dot{\rho}^2}{I^2\rho^{\prime 2}}~.
 \end{equation}
%%%
 Hence, the radial velocity $v^r$ remains finite in
 the limit $t\rightarrow\infty$ irrespective of $n$. From
 Eqs.~\eqref{eq:CT1}-\eqref{eq:CT3}, we have
%%%
\begin{equation}
1-(v^r)^2=\frac{9BI^2}{4\dot{\tau}^2} \label{eq:1-v}
\end{equation}
%%%
and using this relation, the Hubble normalized density $\Omega$ given by
 Eq.~\eqref{eq:g-matter5} becomes
%%%
\begin{equation}
 \Omega=\frac{4\dot{\tau}^2\rho^2}{9R^2\partial_{\rho}R} .
 \end{equation}
%%%
 Applying the approximate form \eqref{eq:appR} of $R$ and the
 asymptotic relation $\dot{\tau}\simeq-\tau$, we have the following
 asymptotic form of the density function 
%%%
\begin{equation}
  \Omega_{\text{asym}}=
\frac{4}{3\{3+(2n+3)F_n\rho^n/|\tau|\}\{1+F_n\rho^n/|\tau|\}}~. \label{eq:ome}
\end{equation}
%%%
Owing to the term $F_n\rho^n/|\tau|$ contained in Eq.~\eqref{eq:ome},
an interesting difference of the asymptotic behavior of $\Omega$
appears according to the choice of $n$ and this will be shown in the
next subsection.

%%%%%%%%%%%%%%%%%%%%%%%%%%%%%%%%%%%%%%%%%%%%%%%%%%%%%%%%%%%%%%%%%%%%%%%
\subsection{Relation between the asymptotic Hubble normalized density
  and the causal structure of the end state}

We rewrite the term $F_n\rho^n/|\tau|$ in Eq.~\eqref{eq:ome} as a
function of $t$ and $r$. In the limit $t\rightarrow\infty$, the metric
functions can be written as $I=I_{0}(r)\exp(-t)$, $J=J_{0}(r)\exp(-t)$
and $R=R_{0}(r)\exp(-t)$. Because the choice of the spatial coordinate
$r$ remains arbitrary within the framework of the separable volume
gauge, the arbitrary function $C(r)$ is included in Eq.~\eqref{eq:J}
for $J$. To remove this ambiguity, we introduce the new spatial
coordinate $\zeta$ by
%%%
\begin{equation}
 -E_{r}^{\ \hat{r}}\frac{d\zeta}{dr}=1~,\label{eq:zeta}
\end{equation}
%%%
where the Hubble normalized variable $E_{r}^{\ \hat{r}}=-I_{0}/J_{0}$
should be regarded as a function of $r$. This specification of the
spatial coordinate $\zeta$ leads to the line element
%%%
\begin{equation}
 ds^{2}=e^{-2t}~[-I_0^2(\zeta)(dt^{2}-d\zeta^{2})+R_0^2(\zeta)\left(d\theta^{2}+
 \sin^{2}\theta d\phi^{2}\right)] ~.\label{eq:LE2}
\end{equation}
%%%

%%%%%%%%%%%%%%%%%%%%%%%%%%%%%%%%%%%%%%%%%%%%%%%%%%%%%%%%%%%%%%%
\subsubsection{The $n \geq 4$ case: Approach to homogeneous dust dynamics
   and black hole formation}

 The $n\geq 4$ case corresponds to the black hole formation and the
 ratio $F_n\rho^n/|\tau|$ approaches zero towards the singularity
 formation.  From Eq.~\eqref{eq:ome}, we have
%%%
\begin{equation}
 \Omega_{\text{asym}}=\frac{4}{9}~.\label{eq:Omega1}
\end{equation}
%%%
In addition, from Eqs.~\eqref{eq:CT1}-\eqref{eq:CT3}, we obtain the
solution for $n\geq4$ as
%%%
\begin{equation}
I_{0}=\frac{Bl^3}{12}\cosh^2\left(\frac{\zeta}{3}\right)~, \qquad 
R_{0}=\frac{Bl^3}{4}\cosh^2\left(\frac{\zeta}{3}\right)
\sinh\left(\frac{\zeta}{3}\right)~\label{eq:n4},
\end{equation}
%%%
where $l$ is an arbitrary constant and
%%%
\begin{equation}
\rho=l\sinh(\zeta/3)\times\exp(-t/3),\qquad
\tau^{2/3}=(Bl^2/4)\cosh^2(\zeta/3)\times\exp(-2t/3) . 
\end{equation}
%%%
Although the metric tensor written by Eqs.~\eqref{eq:n4} is derived as
an asymptotic form in the limit $t\rightarrow\infty$, it is identical
with an exact solution of the Einstein equations describing
spherically symmetric homogeneous dust collapse. This unfamiliar form
of the metric tensor is due to the separable volume gauge; the
timelike congruence parametrized by the coordinates $(t, r)$ (or
$\zeta$) covers only a limited region of the spacetime (see
Fig.~\ref{fig:diagram}). In fact, from Eq.~\eqref{eq:1-v}, the radial
velocity is $v^r=\tanh(\zeta/3)$ and this family of the timelike
curves has the null boundary $|v^r|=1$ at $\zeta=\infty$
($r\rightarrow\infty$).

%%%%%%%%%%%%%%%%%%%%%%%%%%%%%%%%%%%%%%%%%%%%%%%%%%%%%%%%%%%%%%%%%
\subsubsection{The $n=2$ case: Growth of the central density gradient and naked
   singularity formation}

 For $n=2$, the ratio $F_2\rho^2/|\tau|$ grows with the lapse of time
 and the inhomogeneity due to the term $F_n\rho^n$ is
 significantly involved in the central naked singularity
 formation. Under the approximation $F_2\rho^2/|\tau|\gg 1$, we obtain
 the metric functions
%%%
\begin{equation}
I_0=l~,\qquad R_0=l\sinh(\zeta)~\label{eq:n2}
\end{equation}
%%%
and
%%%
\begin{equation}
 F_2^{1/3}\rho^{7/3}=l\sinh\zeta\times \exp(-t),\qquad
\tau^2=(9Bl^2/4)\cosh^2\zeta\times\exp(-2t) .
\end{equation}
%%%
This asymptotic form of $I_{0}$ and $R_{0}$ represents a flat metric
written in the accelerating coordinate system $\{t,\zeta\}$. Using
Eq.~\eqref{eq:1-v}, we have $v^r=\tanh\zeta$. The spacetime region
covered by the accelerating coordinate system also has the null
boundary at $\zeta=\infty$.

The timelike congruence parametrized by $(t, \zeta)$ converges to the
point $\tau=\rho=0$ with the four-velocity $\boldsymbol{u}=
\boldsymbol{e}_{0}$. The spatial distance between the two neighboring
timelike curves $\zeta$ and $\zeta+d\zeta$ changes in proportion to
$\exp(-t)d\zeta$. This change of the reference congruence may be
larger than that of the matter contraction effect which increases the
proper energy density $\tilde{\mu}$. Thus, the effect of matter on the
metric becomes weak and we have the flat metric \eqref{eq:n2} as the
dominant asymptotic behavior. We can find from Eq.~\eqref{eq:ome} that
the Hubble normalized density $\Omega$ asymptotically takes the form
%%%
\begin{equation}
\Omega_{\text{asym}}=
\frac{3B}{7}F_2^{-6/7}l^{2/7}\cosh^2\zeta~(\sinh\zeta)^{-12/7}e^{-2t/7}
\rightarrow 0~. 
\label{eq:Omega2}
\end{equation}
%%%
This exponential decay of $\Omega_{\text{asym}}$ is a remarkable
feature of the asymptotic evolution for the $n=2$ case. As the
contribution of matter to the metric is negligible, one can call the
behavior represented by Eq.~\eqref{eq:Omega2} as the
``vacuum-dominated'' evolution\cite{UgglaC:PRD68:2003}.

It must be noted, however, that the approximation $F_2\rho^2/|\tau|\gg
1$ breaks down near the center $\rho=0$ ($\zeta=0$) which is regular
during $\tau<0$ ($t<\infty$). Let us denote the asymptotic value of
$\Omega$ at the center as $\Omega_0$. Then, we obtain $\Omega_0=4/9$
even for $n=2$. Note that Eq.~\eqref{eq:Omega2} shows a infinite
increase of $\Omega$ in the limit $\zeta\rightarrow 0$. This increase
of $\Omega$ with respect to the spatial coordinate $\zeta$ should be
suppressed if the ratio $F_2\rho^2/|\tau|$ becomes smaller than unity
in the vicinity of the regular center. Unfortunately, it is difficult
to see analytically the smooth decrease of $\Omega$ from the central
value $\Omega_0$. Nevertheless, it is sure that the gradient of
$\Omega$ near the center increases infinitely as $t$ increases. Hence,
the exponential decay \eqref{eq:Omega2} of $\Omega_{\text{asym}}$
should be rather regarded as the growth of the density contrast
between the central region near $\zeta=0$ and the outer region
$\zeta\gg 1$. We expect such a profile of $\Omega_{\text{asym}}$ with
a large gradient with respect to the coordinate $\zeta$ is essential
to the naked singularity formation.

%%%%%%%%%%%%%%%%%%%%%%%%%%%%%%%%%%%%%%%%%%%%%%%%%%%%%%%%%%%%%%%
\subsubsection{The $n =3$ case: Existence of the critical density gradient
   between naked singularity formation and black hole formation\label{sec:n3}}

\begin{figure}[H]
\centering
\includegraphics[width=0.7\linewidth,clip]{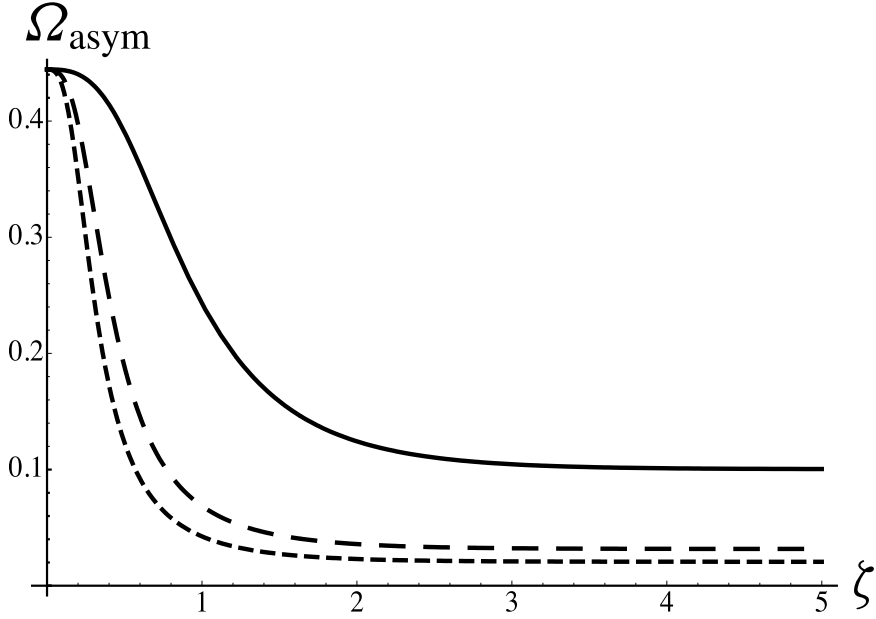}
\caption{The profile of $\Omega_{\text{asym}}$ as a function of the
  spatial coordinate $\zeta$.  Each lines correspond to $b<b_{c}$ (the
  solid line), $b=b_{c}$ (the long-dotted line) and $b>b_{c}$ (the
  short-dotted line). As the value of $b$ increases, the density
  contrast $\delta=\Omega_0/\Omega_{\infty}$ increases. For
  $\delta>\delta_c\approx 15$, the resulting singularity becomes
  naked.}
\label{fig:Omega_asym}
\end{figure}

 In this case, the ratio $F_3\rho^3/|\tau|$ remains finite in the
 limit $t\rightarrow\infty$. As $\tau\sim\rho^3\sim\exp(-t)$
 for this case, by introducing the ratio 
%%%
\begin{equation}
 \label{eq:k}
 k\equiv \frac{F_3\rho^3}{|\tau|}
\end{equation}
%%%
as a function of the spatial coordinate $\zeta$, we obtain from
Eq.~\eqref{eq:ome} the asymptotic profile of $\Omega$ as follows
%%%
\begin{equation}
\Omega_{\text{asym}}(\zeta)=
\frac{4}{9(1+3k)(1+k)}~. 
\label{eq:ome3}
\end{equation} 
%%%
By using Eq.~\eqref{eq:1-v}, the radial velocity $v^r$ can be written
as
%%% 
\begin{equation}
v^r= -\frac{k^{1/3}(1+3k)}{2b^{1/3}(1+k)^{1/3}},\qquad
b=\frac{F_{3}}{B^{3/2}} .
 \label{eq:eta}
\end{equation}
%%%
Then, using Eqs.~\eqref{eq:CT1}-\eqref{eq:CT3}, the equation
determining $k$ is given by
%%%
\begin{equation}
 \frac{dk}{d\zeta} = \frac{k\{1-(v^r)^{2}\}}{v^r}~.
\end{equation}
%%%
Imposing the boundary condition $k=0$ at $\zeta=0$, we obtain the
solution $k=k(\zeta)$ containing the parameter $b$. For $n=3$ case, we
also have $|v^r| \rightarrow 1$ in the limit $\zeta\rightarrow\infty$,
and the value of $k$ remains finite at $\zeta=\infty$. It is clear
from Eq.~\eqref{eq:ome3} that $\Omega_{\text{asym}}$ decreases
monotonically as $k$ increases. Let us denote the values of $k$ and
$\Omega_{\text{asym}}$ in the limit $\zeta\rightarrow\infty$ as
$k_\infty$ and $\Omega_\infty$, respectively. These limiting values
depend on the value of the parameter $b$. The profile of
$\Omega_{\text{asym}}$ as a function of $\zeta$ is shown in
Fig.~\ref{fig:Omega_asym}.
%%%%
%%%%
\noindent
Although $\Omega_{\text{asym}}=4/9$ at the center $\zeta=0$
irrespective of the parameter $b$, the value $\Omega_{\infty}$
decreases as $b$ increases. For $b\gg 1$, we have
$k_{\infty}\simeq2\sqrt{b}/3$ and $\Omega_\infty\simeq 1/3b$. Recall
that $b>b_{c}$ gives the condition for the naked singularity
formation. The numerical calculation shows that this inequality
corresponds to $\Omega_\infty<\Omega_c\simeq 0.03$. Using the central
value $\Omega_0=4/9$ of $\Omega_{\text{asym}}$, the density contrast
defined by the ratio $\delta=\Omega_0/\Omega_\infty$ has a critical
value $\delta_c$: $\delta<\delta_c$ corresponds to the black hole
formation and $\delta>\delta_c$ corresponds to the naked singularity
formation. For $\delta\gg \delta_c$, the density contrast becomes very
large and this leads to the naked singularity formation. This behavior
is consistent with the asymptotic profile of $\Omega$ discussed in the
case $n=2$.%
\footnote{Although the naked singularity formation occurs both for
  $n=2$ and for $n=3$ with the parameter $b>b_c$, its strength is
  known to be different in the sense of Tipler
  \cite{TiplerFJ:PL64:1977}. Namely, the arising central singularity
  is called gravitationally weak for the former case but strong for
  the latter one \cite{JhinganS:CQG13:1996}. This difference of the
  strength can also be interpreted from the viewpoint of the
  asymptotic behavior of $\Omega$, which claims that the
  ``vacuum-dominated'' evolution under the condition
  $\Omega\rightarrow 0$ around the central region results in the
  formation of ``weak'' naked singularity.}

We finally comment on the self-similar evolution in dust collapse in
terms of the Hubble normalized variables. The spherically symmetric
self-similar solutions are defined as solutions of the Einstein
equations reduced to a set of the ordinary differential equations with
respect to a dimensionless variable (see \cite{CarrBJ:CQG16:1999} for
its details). For $n=3$, the Hubble normalized density $\Omega$, as
well as the other Hubble normalized variables $(v^{r}, \lambda^{r},
A^{r}, E_{r}^{\ \hat{r}}, E_{\theta}^{\ \hat{\theta}})$ is given as
the function of the single coordinate $r$ in the limit
$t\rightarrow\infty$. Although we have derived the asymptotic behavior
of $v^r$ and $\Omega$ through the coordinate transformations
\eqref{eq:CT1}-\eqref{eq:CT3}, it is possible to obtain all the
asymptotic forms of these variables directly from an autonomous system
of the evolution equations \eqref{eq:EE1}-\eqref{eq:EE6} and the
constraints \eqref{eq:C1}-\eqref{eq:C3} for $\gamma=1$. As
$\Sigma^{rr}$ and $q$ approach zero in the limit $t\rightarrow\infty$,
if these two variables in addition to variables with partial
derivatives with respect to $t$ are neglected in the equations, we
arrive at a closed system of the ordinary differential equations for
$(\Omega, v^{r}, \lambda^{r}, A^{r}, E_{\theta}^{\ \theta})$ using the
dimensionless coordinate $\zeta$(see the next section for the
details). This closed system will be equivalent to the Einstein
equations with the requirement of self-similarity.%
\footnote{In fact, we can check that the function
  $\Omega_{\text{asym}}$ given by Eq.~\eqref{eq:ome3} has the same
  form as Eq.~(3.13) in \cite{CarrBJ:PRD62:2000} except for the
  numerical factor.} %
The result obtained in this subsection means that such a self-similar
behavior should develop at the final stage $t\rightarrow\infty$, even
if the initial distribution function $F(\rho)$ given by
Eq.~\eqref{eq:F} contains the higher order inhomogeneous terms with
$\rho^{n'}$ for $n'\geq 4$ in addition to the term $F_3\rho^3$.

%%%%%%%%%%%%%%%%%%%%%%%%%%%%%%%%%%%%%%%%%%%%%%%%%%%%%%
\subsection{Condition for the naked singularity formation}

Combining the result of our analysis for the asymptotic behavior of
the Hubble normalized density function $\Omega$ in the LTB solution,
we propose the following conjecture for the condition of the naked
singularity formation:

\textit{The development of the large spatial gradient or density
  contrast in the asymptotic profile of the Hubble normalized density
  parameter $\Omega$ in the separable volume gauge gives a sufficient
  condition of the naked singularity formation.}

We have confirmed this conjecture holds for the dust case. In the next
section, we examine whether this conjecture can be extended to the
gravitational collapse of perfect fluid with pressure.

%%%%%%%%%%%%%%%%%%%%%%%%%%%%%%%%%%%%%%%%%%%%%%%%%%%%%%%%%%%%%%%%%%%%%%%%%%%
\section{Effect of pressure on the Hubble normalized density  \label{sec:PF}}

In the previous section, we have discussed the asymptotic behavior of
the Hubble normalized variables through the coordinate transformation
from the LTB solution of dust collapse. Unfortunately, there does not
exist such a generic analytical solution of perfect fluid collapse for
the equation of state \eqref{eq:state} with $1< \gamma \leq 2$. We
therefore have to analyze directly the asymptotic solutions of an
autonomous system of the evolution equations
\eqref{eq:EE1}-\eqref{eq:EE6} and the constraints
\eqref{eq:C1}-\eqref{eq:C3}.

As was schematically shown in Fig.~\ref{fig:diagram}, under the
separable volume gauge \eqref{eq:J} for the line element
\eqref{eq:LE}, we can assume the timelike congruence parametrized by
the coordinates $t$ and $r$ converges to the central singularity
arising at $t=\infty$ and the possible asymptotic time dependence can
be assumed to be $I\sim J\sim R\sim\exp(-t)$. The Hubble normalized
variables $q$ and $\Sigma_{r}^{\ r}$ then go to zero in the limit
$t\rightarrow\infty$, while the other variables $(E_{r}^{\ \hat{r}},
E_{\theta}^{\ \hat{\theta}}, A^{r}, \lambda^{r}, \Omega, v^{r})$ may
become finite dependent only on $r$. We express their asymptotic forms
as $(E, A, \lambda, \Omega, v)$ without the indices $r$ and $\theta$,
and also introduce the coordinate $\zeta$ given by
Eq.~\eqref{eq:zeta}. It is easy to obtain $\lambda=-3\gamma\Omega
v/2G$ from Eq.~\eqref{eq:C2} and $dE/d\zeta=-(A+\lambda)E$ from
Eq.~\eqref{eq:C3}. The assumptions for the timelike congruence and the
possible asymptotic time dependence of $(I, J, R)$ allow us to arrive
at the closed set of equations for $(A, \Omega, v)$ as follows
%%%
\begin{align}
 &\left\{(\gamma -1)v^{2}+1\right\}\left\{v^{2}
-(\gamma -1)\right\}\frac{d\Omega}{d\zeta}= \notag \\
 &\qquad 2\gamma \Omega v^{2}A\left\{2\gamma -3-(\gamma -1)v^{2}\right\}
-\frac{3}{2}\gamma(\gamma-2)G^{-1}\Omega^{2}v\left\{1-(\gamma-1)v^{2}\right\}
(1-v^{2})\notag\\
 &\qquad\qquad  -2\Omega v\left\{(3\gamma -4)(\gamma -1)-(2\gamma ^{2}-5\gamma
   +4)v^{2}\right\}~,\label{eq:sde1}\\
 &\frac{\gamma\left\{(\gamma -1)v^{2}+1\right\}\left\{v^{2}
     -(\gamma -1)\right\}}{1-v^{2}}\frac{dv}{d\zeta}=\notag\\
 &\qquad 2\gamma (\gamma -1)GvA+\frac{3}{2}\gamma ^{2}(\gamma -2)\Omega
 v^{2}-G\left\{(\gamma -1)(3\gamma -2)+(\gamma
   -2)v^{2}\right\}~,\label{eq:sde2}\\
 &\frac{dA}{d\zeta}=3\gamma G^{-1}\Omega
 vA-\frac{3}{2}G^{-1}(\gamma-1+v^{2}) 
\Omega -\frac{3}{2}\Omega -A^{2}+1~.\label{eq:sde3}
\end{align}
%%%
From Eq.~\eqref{eq:sde2}, we can find that there exists the null
boundary $|v|=1$ of the timelike congruence in the limit
$\zeta\rightarrow\infty$ just in the same way as dust collapse.

It should be noted that this set of equations for $(A, \Omega, v)$ is
derived under the assumption $\Omega\neq 0$. If we consider the case
$\Omega\rightarrow 0$ in the limit $t\rightarrow\infty$, we must use
the equation
%%%
\begin{equation}
E^2-A^2+1=0
\end{equation}
%%%
instead of Eqs.~\eqref{eq:sde1} and \eqref{eq:sde2}. The similar
equation can be obtained for the case $n=2$ of the dust collapse. In
fact, under the requirement $\Omega=\lambda=0$, we have the solution
$E=-1/\sinh\zeta, A=\coth\zeta$ which corresponds to the flat metric
given by \eqref{eq:n2}. It was shown in the previous section that this
``vacuum-dominated'' evolution should break down in the vicinity of
the regular center due to the large spatial gradient of
$\Omega_{\text{asym}}$. For the perfect fluid collapse, the effect of
pressure should become important at least in the subsonic region
$v^2<\gamma-1$ near the center and works to suppress the development
of the density contrast.  Nevertheless, it is plausible that the
``vacuum-dominated'' evolution leading to the decay $\Omega\rightarrow
0$ is allowed in the supersonic region $v^2>\gamma-1$, where pressure
becomes ineffective. This means that a large gradient of
$\Omega_{\text{asym}}$ generated in the supersonic region is in favor
of the naked singularity formation. Unfortunately, it is very
difficult to study in details the perfect fluid collapse allowing the
decay of $\Omega$ with the lapse of time in the supersonic region and
to check the validity of our expectation. Therefore, in the following,
our discussion is restricted to the case that $\Omega$ remains nonzero
in the limit $t\rightarrow\infty$ in $0\leq\zeta<\infty$, and
Eqs.~\eqref{eq:sde1}-\eqref{eq:sde3} can be applied.

The set of the ordinary differential equations
\eqref{eq:sde1}-\eqref{eq:sde3} with respect to the nondimensional
coordinate $\zeta$ describes the self-similar behavior of the Hubble
normalized variables such as $\Omega$. Although in this paper, we
assume that they become asymptotically dominant equations as a result
of gravitational collapse started from general non self-similar
initial conditions, they are equivalent to the exact self-similar
Einstein equations which have been extensively studied in the comoving
coordinate system. We can see the existence of a singular point of the
differential equations at $v^2=\gamma-1$ where the coefficients of the
derivatives $d\Omega/d\zeta$ and $dv/d\zeta$ in Eqs.~\eqref{eq:sde1}
and \eqref{eq:sde2} vanish. This corresponds to the sonic point where
the fluid velocity relative to a tangent surface of a homothetic
Killing vector equals to the sound speed \cite{OriA:PRD42:1990,
  FoglizzoT:PRD48:1993}, and the self-similar solutions may become
singular.

We consider the self-similar solutions regular in $0\leq\zeta<\infty$
including the sonic point. The simplest example is the so-called flat
Friedmann solution with $\gamma>1$. It is well known that the arising
singularity becomes spacelike in the homogeneous perfect fluid
collapse. For this solution, the asymptotic density function is given
by
%%%
\begin{equation}
 \Omega_{\text{asym}}
 = \frac{4}{9\gamma^{2}}\left[\frac{1+\gamma\sinh^2\{(3\gamma-2)\zeta/3\gamma\}}
 {1+\sinh^2\{(3\gamma-2)\zeta/3\gamma\}}\right]~. \label{eq:OmegaF}
\end{equation} 
%%%
The $\zeta$ dependence of this function is shown in Fig.~3.  This form
can be obtained through the coordinate transformation from the
comoving coordinate system to the coordinate system
$\{t,r,\theta,\phi\}$ as was performed in Sec.~\ref{sec:LTB}. Of
course, it is also possible to derive this form directly from
Eqs.~\eqref{eq:sde1}-\eqref{eq:sde3} with
$v^r=-\tanh\{(3\gamma-2)\zeta/3\gamma\}$. This solution can be
interpreted as an extension of the case $n\geq 4$ of the dust
collapse, for which we have $\Omega_{\text{asym}}=4/9$. By virtue of
the effect of pressure, we have the density contrast
$\delta=\Omega_0/\Omega_\infty=1/\gamma<1$ in this homogeneous perfect
fluid collapse.
%%%
\begin{figure}[H]
  \centering
  \includegraphics[width=0.7\linewidth,clip]{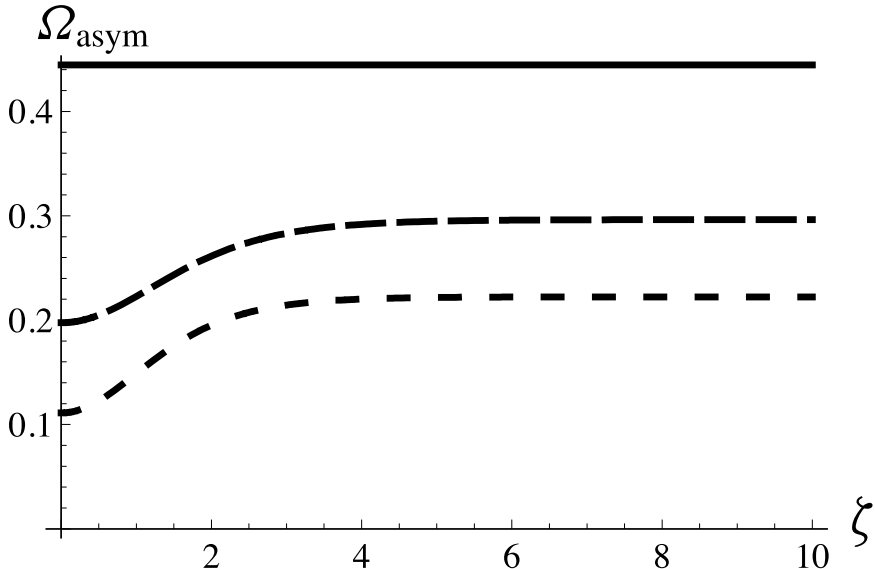}
  \caption{The spatial profile of $\Omega_{\text{asym}}$ for the flat
    Friedmann solution with perfect fluid. Each lines correspond to $\gamma=1$
    (the solid line), $\gamma=3/2$ (the dotted-line) and $\gamma=2$
    (the short-dotted line).}
\end{figure}
%%%
\noindent
As the density contrast or the spatial gradient is not so  large,
the evolution cannot become ``vacuum-dominated'' and the resulting
singularity is not naked. This example supports our conjecture.

The effect of pressure becomes clearer if we consider the general
relativistic Larson-Penston (GRLP)
solution\cite{OriA:PRD42:1990}. This is the numerically obtained
self-similar solution which describes monotonic collapse of an
inhomogeneous perfect fluid with the regularity imposed at the sonic
point. This self-similar solution may correspond to the case $n=3$ of
the dust collapse provided that $\gamma$ is treated as a free
parameter instead of $b$. The central singularity of this solution
becomes naked for $\gamma < \gamma_c \simeq 1.0105$ (weak pressure),
while it is hidden behind a horizon for $\gamma>\gamma_c$ (strong
pressure).

\begin{figure}[H]
\centering
\includegraphics[width=0.7\linewidth,clip]{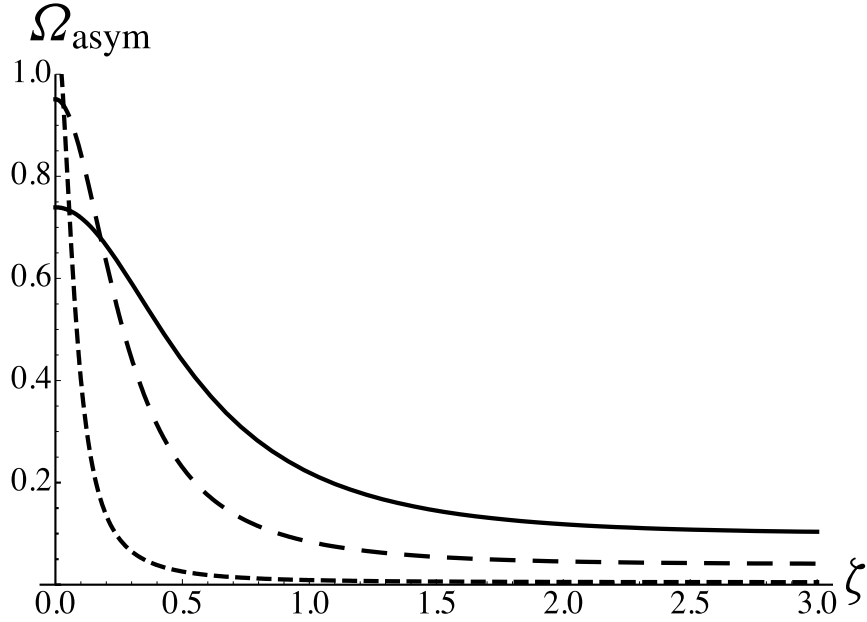}
\caption{The profile of the Hubble normalized density
  $\Omega_{\text{asym}}$ corresponding to the GRLP solution. Each
  lines correspond to $\gamma>\gamma_c$ (the solid line),
  $\gamma=\gamma_c$ (the long-dotted line), $\gamma<\gamma_c$ (the
  short-dotted line). As the value of $\gamma$ decreases, the density
  contrast $\delta=\Omega_0/\Omega_\infty$ increases. For
  $\delta>\delta_c\approx 24$, the resulting singularity becomes
  naked.}
\label{fig:Omega_GRLP}
\end{figure}

To obtain $\Omega_{\text{asym}}$ corresponding to the GRLP solution,
we numerically solve the ordinary differential equations
\eqref{eq:sde1}-\eqref{eq:sde3} by requiring the regularity of the
solution both at the center $\zeta=0$ and the sonic point
$v^2=\gamma-1$. The result is shown in Fig~\ref{fig:Omega_GRLP}.
%%%
%%%
\noindent
It is clear from Fig.~\ref{fig:Omega_GRLP} that $\Omega_{\text{asym}}$
decreases monotonically as $\zeta$ increases, just in the same way as
the case $n=3$ of the dust model (see
Fig.~\ref{fig:Omega_asym}). Further, we note that the density contrast
$\delta=\Omega_0/\Omega_\infty$ in this self-similar perfect fluid
collapse increases with the decrease of $\gamma$. In particular, the
critical value of $\gamma$ for the naked singularity formation is
$\gamma = \gamma_c \simeq 1.0105$ and using this value, we can roughly
estimate the corresponding critical density contrast to be
$\delta_c=\Omega_0/\Omega_\infty\simeq 24\gg 1$ ($\Omega_\infty\simeq
0.04$). Recall that for the case $n=3$ of dust collapse, we have the
critical value $\delta_c=\Omega_0/\Omega_\infty\simeq 15\gg
1$. Although the critical value of the density contrast will slightly
depend on the property of collapsing matter, the asymptotic analysis
of the perfect fluid collapse also supports our conjecture that the
very steep change in the asymptotic spatial profile of $\Omega$
characterizes the type of the arising singularity.

The asymptotic behavior of the Hubble normalized variable with the
separable volume gauge becomes self-similar provided that the
asymptotic value of $\Omega$ is not zero. This may be interesting in
relation to the so-called self-similar
hypothesis\cite{CarrBJ:GRG37:2005} which asserts that a self-similar
behavior should be dominant near the dense central region at the final
stage of collapse starting from general initial conditions.

%%%%%%%%%%%%%%%%%%%%%%%%%%%%%%%%%%%%%%%%%%%%%%%%%%%%%%%%
\section{Summary and discussion \label{sec:sum}}

We have studied the asymptotic dynamics of the naked singularity
formation in spherically symmetric gravitational collapse of perfect
fluid.  We first have examined inhomogeneous dust gravitational
collapse described by the marginally bound LTB solution.  As our main
result, we have revealed the different asymptotic behavior of the
Hubble normalized density $\Omega$ depending on the type of the
initial inhomogeneity. By comparing to the known causal structure of
singularity arising in the dust collapse, we arrive at the important
conjecture that the very steep decrease in the asymptotic spatial
profile of $\Omega$ is the characteristic of the naked singularity
formation.  The validity of this conjecture has been also supported in
the perfect fluid collapse with pressure.

Let us remark on the ``vacuum-dominated'' case of the dust collapse
(the case $n=2$), for which $\Omega$ goes to zero in the limit
$t\rightarrow\infty$ and the arising singularity becomes
naked. Because the central value $\Omega_0$ should remain nonzero, the
vacuum-dominated evolution leading to the naked singularity formation
has been interpreted as the development of the steep spatial gradient
in the profile of $\Omega$ with the lapse of time. For the dust
collapse, as was mentioned in Sec.~\ref{sec:LTB}, the asymptotic
behavior $\Omega\rightarrow 0$ with the naked singularity formation is
rather generic. Unfortunately, for perfect fluid collapse with
pressure, the dynamical evolution from an initial state to the
vacuum-dominated state $\Omega\rightarrow 0$ is not confirmed. One may
find that $\Omega$ approaches zero only in the outer supersonic region
$|v^r|>\sqrt{\gamma-1}$. The self-similar behavior may be then
dominant in the inner subsonic region. To reveal the existence of the
vacuum-dominated evolution and the structural change from the inner
region to the outer one remains as an important problem to be
investigated, especially, in relation to the criterion for the naked
singularity formation.

If no symmetry is assumed in gravitational collapse, one can assert
that the generic singularity becomes spacelike from the so-called BKL
conjecture \cite{BelinskiiVA:AP19:1970,BelinskiiVA:AP31:1982} that
states the Einstein equation becomes local near the singularity. In
terms of the 1+3 orthonormal frame formalism considered here, the BKL
conjecture has been reformulated in \cite{UgglaC:PRD68:2003} and
applied to various types of collapsing matter (e.g., the perfect fluid
with $\gamma=2$ \cite{CurtisJ:PRD72:2005}). Although the BKL
conjecture has been supported even in the vacuum-dominated case, the
arising singularity is known to be null (see
\cite{ChristodoulouD:CMP93:1984, SinghTP:CQG13:1996}) for spherically
symmetric dust collapse if $\Omega\rightarrow0$. In this paper, we
have claimed that the naked singularity is originated from the steep
gradient in the profile of $\Omega$. For nonspherically symmetric
collapse, of course, the freedom of gravitational waves becomes
important in the vacuum-dominated case and the long-wavelength modes
will be crucial to suppress such a steep density gradient and avoid
the naked singularity formation. The BKL conjecture on the causal
structure of singularity is based on the assumption that the spatial
gradient of dynamical variables becomes negligible at the final stage
of collapse. This may not be valid for some models of gravitational
collapse including spherically symmetric models.

%\bibliography{relativity}

\end{document}